\documentclass[12pt]{iopart}

\usepackage{iopams}

\expandafter\let\csname equation*\endcsname\relax

\expandafter\let\csname endequation*\endcsname\relax

\usepackage{amsmath}

\usepackage{amstext}
\usepackage{graphicx}
\usepackage{mathbbol}
\usepackage{color}

\def\Tr{\mbox{Tr}}

\begin{document}

\title[]{Feedback controlled heat transport in quantum devices: Theory and solid state experimental proposal}
\author{Michele Campisi}
\address{NEST, Scuola Normale Superiore \& Istituto Nanoscienze-CNR, I-56126 Pisa, Italy}
\ead{michele.campisi@sns.it}
\author{Jukka Pekola}
\address{Low Temperature Laboratory, Department of Applied Physics, Aalto University School of Science, 00076 AALTO, Finland}
\ead{pekola@ltl.tkk.fi}
\author{Rosario Fazio}
\address{ICTP, Strada Costiera 11, Trieste 34151, Italy}
\address{NEST, Scuola Normale Superiore \& Istituto Nanoscienze-CNR, I-56126 Pisa, Italy}
\ead{rosario.fazio@sns.it}

\begin{abstract}
A theory of feedback controlled heat transport in quantum systems is presented. It is based on modelling heat engines as driven multipartite systems subject to projective quantum measurements and measurement-conditioned unitary evolutions. The theory unifies various results presented in the previous literature. Feedback control breaks time reversal invariance. This  in turn results in the fluctuation relation not being obeyed. Its restoration occurs by an appropriate accounting of the information gain and information use via measurements and feedback. We further illustrate an experimental proposal for the realisation of a Maxwell demon using superconducting circuits and single photon on-chip calorimetry. A two level qubit acts as a trapdoor which, conditioned on its state is coupled to either a hot resistor or a cold one. The feedback mechanism alters the temperatures felt by the qubit and can result in an effective inversion of temperature gradient, where heat flows from cold to hot thanks to information gain and use.
\end{abstract}
\maketitle

\section{Introduction}
In a famous thought experiment Maxwell envisioned a method for apparently defying the second law of thermodynamics
by means of a feedback control mechanism \cite{Maruyama09RMP81}. Maxwell's idea is based on a malicious demon, an intelligent being that is able to observe the microscopic dynamics of a system, and acts on it so as to steer it toward defying the second law. In one of Maxwell's original concepts, the system is a container with two chambers, containing respectively a hot gas and a cold gas. The two chambers are separated by a wall presenting a trap-door which the demon can open and close at will. The demon observes the erratic motion of the gas particle and when sees a particle of the cold chamber approach the trap-door with sufficiently high velocity, she/he swiftly opens the door as to let the particle go through and closes it immediately afterwards. In this way, particle after particle, heat flows from the cold chamber to the hot chamber in contradiction with the second law.

Advance in nanotechnology has made the possibility of bringing Maxwell demons and similar devices from the realm of thought experiments to the realm of real experiments \cite{Toyabe10NP6,Koski14PRL113,Koski14PNAS111,Berut12Nature483}. Both theoretical and experimental studies so far have focused mainly on situations where feedback control is operated as a measurement-conditioned driving on some working substance (classical or quantum) coupled to a single temperature, so as to withdraw energy from the latter in contradiction with the second law as formulated by Kelvin. Interesting realistic proposals have appeared in Refs. \cite{Strasberg13PRL110,Bergli13PRE88}. Situations where heat flows between different temperature reservoirs is controlled, however have not been addressed so far, neither theoretically nor experimentally.
The main motivation of the present work is that of filling that gap. In the following we shall present the general theory of feedback controlled heat transport in quantum devices, and shall describe a possible experimental realisation thereof. 

The theory presented here builds on previous works concerning  fluctuation relations in presence of measurements without feedback \cite{Campisi10PRL105,Campisi11PRE83} and with feedback \cite{Morikuni11JSP143}, combined with an inclusive approach where quantum heat engines are seen as mechanically driven multipartite systems starting in a multi-temperature initial state \cite{Jarzynski99JSM96,Campisi14JPA47,Campisi15NJP17,Campisi16JPA49a}.
Reference \cite{Morikuni11JSP143} reported on the theory of a one-measurement based feedback control on a quantum working-substance prepared by contact with a single bath. That formalism is here extended to the case of many heat baths, and also repeated measurements, to allow for the study of continuous feedback control of heat flow in a multi reservoir scenario. Previous work concerning repeated measurements appeared in Refs \cite{Horowitz10PRE82} for classical systems in contact with a single bath. Fluctuation relations need to be modified by a mutual information term, which we shall explicitely provide.

Our experimental proposal is based on the fast developing advancements in experimental solid state low temperature techniques: in particular the calorimetric measurement scheme that has been put forward by one of us and co-workers \cite{Pekola10PRL105,Gasparinetti15PRAPP3}. As proven by some recent theoretical proposals \cite{Campisi15NJP17,Karimi16PRB94} the method opens up a new avenue for the practical management of heat and work on a chip by means of superconducting devices, particularly superconducting qubits. Here we illustrate the possible implementation of very simple feedback controlled heat transport where the trapdoor is realised by a superconducting qubit whose coupling with two resistors at different temperatures is controlled based on the outcomes of continuous calorimetric monitoring of the resistors themselves.

\begin{figure}[b]
\includegraphics[width=\linewidth]{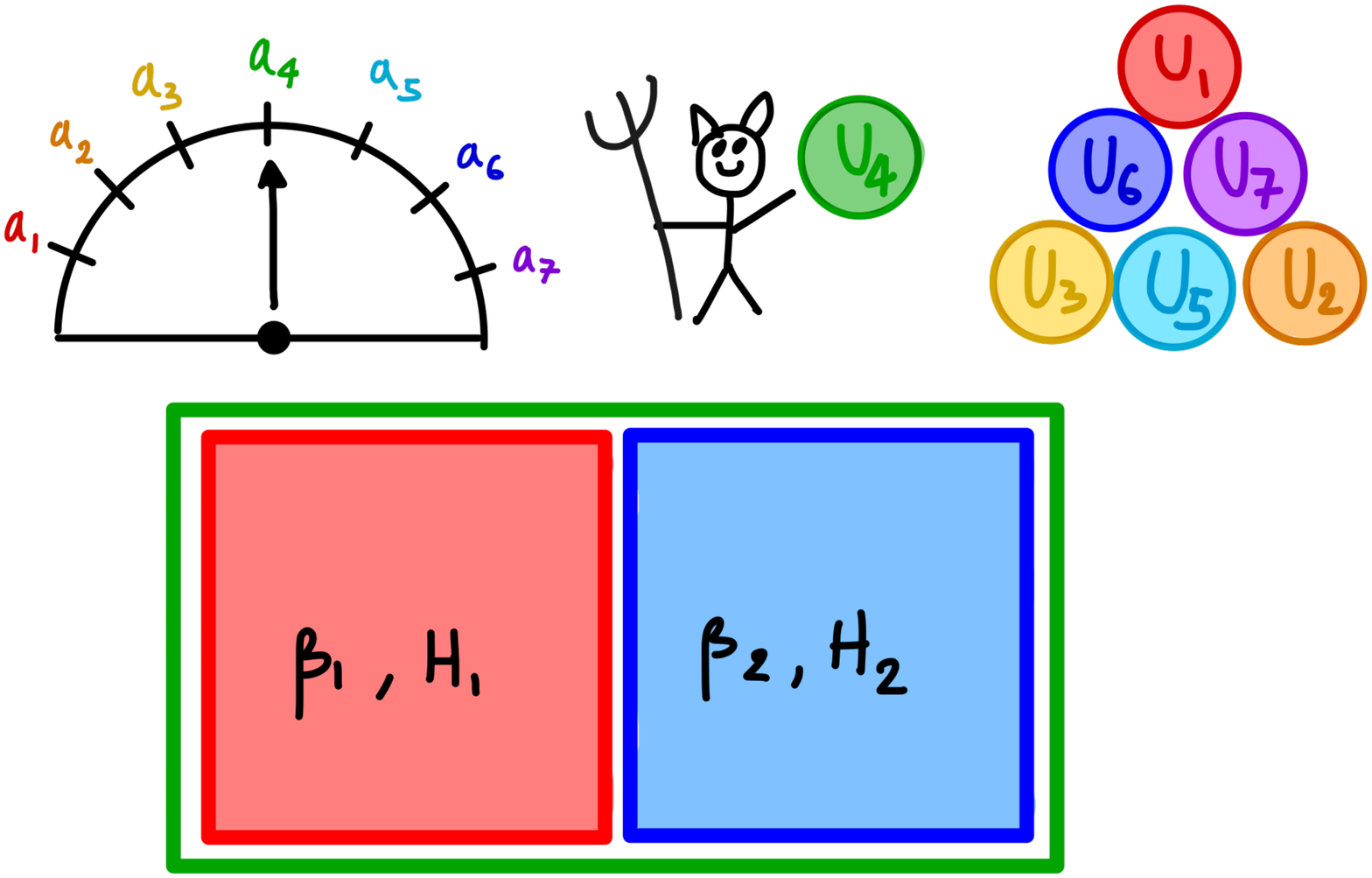}
\caption{Feedback controlled heat transport. A bi-partite system starting in a two temperature Gibbs state is observed by a Demon, who measures an observable $A$. Depending on the outcome $a_j$ of the measurement the demon applies a quantum gate $U_j$ to the bi-partite system with the aim of beating the second law. Each partition is composed of a heat reservoir and possibly one part of a working substance. The whole system evolves with unitaries interrupted by projections.}
\label{fig:1}
\end{figure}

\section{Theory}
Following \cite{Campisi16JPA49a} we model a generic heat transport/heat engine scenario as a driven multi-partite system starting in the factorised state, see Figure \ref{fig:1}
\begin{align}
\rho_0 =  \bigotimes_{l} \frac{e^{-\beta_l H_l}}{Z_l} \label{eq:eq-initial1}
\end{align}
where $H_i$ is the Hamiltonian of each partition including a heat bath and possibly a portion of the working substance, and $Z_i$ is the corresponding partition function \cite{Campisi16JPA49a}.
Let the total Hamiltonian be 
\begin{align}
H(t) = \sum_l H_l+ V(t)
\end{align}
where  $V(t)$ is an interaction term that is switched on for the time interval $t \in [0,\tau]$ over which the system is monitored. We assume that at times $t_1<t_2< \dots t_K$ some observable $A$ is measured thus causing the wave function describing the compound to collapse onto the subspace spanned by the eigenvectors belonging to the measured eigenvalue $a_j$. Following  \cite{Morikuni11JSP143} we shall assume that there can be a measurement error where the eigenvalue $a_{k}$ is recorded instead of the actual eigenvalue $a_j$. This is assumed to happen with probability $\varepsilon [k|j]$. 
The choice of the interaction $V(t)$ in the interval $(t_{i},t_{i+1})$ is dictated by the sequence of recorded eigenvalues, or more simply the recorded sequence  $\{k_1,k_2,... k_{i}\}=\mathbf{k}_i$,  that is for $t_i<t<t_{i+1}$ $V(t)=V_{\mathbf{k}_i}(t)$. The corresponding unitary operator describing the evolution in the time span $t_i<t<t_{i+1}$ is $U_{\mathbf{k}_i}= \overleftarrow{\exp}\left[\int_{t_i}^{t_{i+1}}ds\,  H_{\mathbf{k}_i}(s) \right]$ where $\overleftarrow{\exp}$ denotes time ordered exponential  and $H_{\mathbf{k}_i}(t)=  \sum_l H_l+ V_{\mathbf{k}_i(t)}$. We shall denote the un-conditioned evolution operator from time $t=0$ to the time of the first measurement $t= t_1$ as $U_0$.
Note that the sequence of recorded labels $\mathbf{k}_j$ generally differs from the sequence of labels $\{j_1,j_2,... j_{i}\}=\mathbf{j}_i$ specifying in which subspace the system state was actually projected at the measurement times $t_1, t_2, \dots t_i$.
As customary in the context of the fluctuation theorem we shall assume that besides the intermediate measurements of $A$, all $H_l$'s are measured at times $t=0,t=\tau$ giving the eigenvalues $E_n^l$, $E_m^l$ respectively.

The quantity of primary interest is the probability $p(m,\mathbf{k},\mathbf{j},n)$ that $n$ is obtained in the first energy measurement, the sequence $\mathbf{j}$ is realised, the sequence $\mathbf{k}$ is recorded and $m$ is obtained in the final energy measurement. Here we have introduced the simplified notations $\mathbf{j}=\mathbf{j}_K$, $\mathbf{k}=\mathbf{k}_K$. The explicit expression of $p(m,\mathbf{k},\mathbf{j},n)$ is:
\begin{align}
\label{eq:pnjkm}
p(m,\mathbf{k},\mathbf{j},n) &=
\Tr\,  P_m A_{\mathbf{k},\mathbf{j}} U_0  P_n U_0^\dagger A_{\mathbf{k},\mathbf{j}}^\dagger P_m p^0_n \\
A_{\mathbf{k},\mathbf{j}}&=\overleftarrow{\Pi}_i\left(U_{\mathbf{k}_i}\pi_{j_i}\sqrt{\varepsilon[k_i | j_i]}\right)
\end{align}
where $p^0_n= \Pi_l e^{- \beta_l E_n^l}/Z_l$ denotes the probability of obtaining the eigenvalue $E_n= \sum_l E_n^l$ in the first measurement; $P_n$ denotes the corresponding projector; $\pi_j$ denotes the projector onto the subspace belonging to the eigenvalue $a_j$ of $A$; the symbol $\overleftarrow{\Pi}_i$ denotes $i$-ordered product, that is,  
$\overleftarrow{\Pi}_i(U_{\mathbf{k}_i}\pi_{j_i}\varepsilon[k_i | j_i])= U_{\mathbf{k}_K}\pi_{j_K}\sqrt{\varepsilon[j_K | k_K]} \cdots U_{\mathbf{k}_2}\pi_{j_2}\sqrt{\varepsilon[j_2 | k_2]} U_{\mathbf{k}_1}\pi_{j_1}\sqrt{\varepsilon[j_1| k_1]}$.

Let $\Delta E_l = E_m^l-E_n^l$ be the energy change in the partition $l$ observed in a single realisation of the feedback driven protocol. Using the cyclic property of the trace and completeness $\sum P_n=\mathbb{1}$, we obtain the following:
\begin{align}
\label{eq:FT1}
\langle e^{-\sum_l \beta_l \Delta E_l} \rangle = \gamma = \sum_{\mathbf{j},\mathbf{k}} \Tr A_{\mathbf{k},\mathbf{j}}^\dagger \rho_0 A_{\mathbf{k},\mathbf{j}}
\end{align}
The proof is reported in the appendix. This relation extends the result presented in Ref. \cite{Morikuni11JSP143} to the case of multipartite system with initial multi temperature state, and to repeated measurements.\footnote{For simplicity we restricted to the case of cyclic $H(t)$. The extension to non-cyclic case is straightforward.} The quantity $\Tr A_{\mathbf{k},\mathbf{j}}^\dagger \rho_0 A_{\mathbf{k},\mathbf{j}}$
 represents the probability that the sequences $\mathbf{j}^\dagger= \{j_K, ...j_2,j_1\}$, $\mathbf{k}^\dagger= \{k_K, ... k_2,k_1\}$,  are realised under the backward evolution specified by the adjoint Kraus operators $A_{\mathbf{k},\mathbf{j}}^\dagger$. The total probability $\gamma$ does not generally add to one.
The reason for that is that the $i$-th evolution $U_{\mathbf{k}_i}^\dagger$ occurs before the the $i$-th eigenvalue $j_i$ is realised in the backward map. The feedback loop is evidently not time-reversal symmetric, and such lack of reversibility breaks the fluctuation theorem $\langle e^{\sum_l \beta_l \Delta E_l} \rangle = 1$ which in fact is a manifestation of time-reversal symmetry \cite{Campisi11RMP83}. This is reflected by the fact that the quantum channel specified by the Kraus operators $A_{\mathbf{k},\mathbf{j}}$ is generally not \emph{unital}.\footnote{We recall that a quantum channel specified by Kraus operators $M_i$, $\rho \rightarrow  \sum_i  M_i \rho M_i^\dagger$ that is trace preserving $\sum M_i^\dagger M_i  = \mathbb{1}$, is unital when it maps the identity into itself $\sum M_i M_i^\dagger = \mathbb{1}$.} The adjoint of a non-unital quantum channel is not trace preserving. In the case of feedback control the quantum channel $\sum_{\mathbf{j},\mathbf{k}} A_{\mathbf{k},\mathbf{j}}\rho_0 A_{\mathbf{k},\mathbf{j}}^\dagger$ is generally not unital, as a consequence its adjoint is generally not trace preserving, hence we have generally $\gamma \neq 1$. Lack of unitality generally reflects lack of time-reversal symmetry. Examples are thermalisation maps, namely maps that have a thermal state (not the identity) as a fixed point. Physically these are realised by means of weak contact of a system with a thermal bath, leading to irreversible dynamics. Likewise feedback control breaks the symmetry. This observation reveals some analogy between feedback control and dissipative dynamics.

Before proceeding let us comment briefly on the origin of lack of unitality in feedback controlled systems, in order to gain insight in the issue. For simplicity let us consider the case of a single measurement $K=1$. Let us begin by noticing that the quantum channel specified by the $A_{k,j}$ is trace preserving. We have $\Tr \sum_{k,j} A_{k,j} \rho A_{k,j}^\dagger  = \sum_{k,j}  \varepsilon[k|j]  \Tr \, U_{k}\pi_j \rho \pi_j U_k ^\dagger  = \sum_{k,j}  \varepsilon[k|j]  \Tr\,  \pi_j \rho \pi_j = \sum_{j}  \Tr\,  \pi_j \rho \pi_j =\Tr \, \rho$,
where we have used the cyclic property of the trace, unitarity $U_{k}^\dagger U_k=\mathbb{1}$, idempotence $\pi_j \pi_j = \pi_j$, normalisation $\sum_{k} \varepsilon[k|j] = 1$, and completeness $\sum \pi_j = \mathbb{1}$. Let us now turn to unitality.
 We have $\sum_{k,j} A_{k,j}A_{k,j}^\dagger  = \sum_{k,j} \varepsilon[k|j] U_{k}\pi_j U_k^\dagger$. If the evolution $U_k$ did not dependent on $k$, that is $U_k=\bar{U}$ was chosen regardless of the recorded value $k$ (e.g, $\bar{U}$ is pre-specified or is completely random), one could perform the sum over $k$ using $\sum_k \varepsilon[k|j]=1$ and then use $\sum \pi_j = \mathbb{1}$ to conclude the map is unital. Feedback, implying explicit dependence on $k$ of $U_k$ breaks unitality. Unitality would occur  also in the case when $\varepsilon[k|j]$ does not depend on $j$, meaning the measurement outcome $k$ is completely random and has no correlation with the actual state $j$. In sum if the feedback control measurement is off, either because one decides not to use the information gathered in the measurement, or because the measurement gathers no information in the first place, unitality is recovered, and the fluctuation theorem is restored. This result is in agreement with the established fact that projective measurements without feedback control do not alter the validity of the fluctuation theorem \cite{Campisi10PRL105,Campisi09PRE80,Watanabe14PRE89}. Here we have further learned that noise, i.e. choosing the $U$'s between the measurements completely randomly, also does not affect the integral fluctuation relation.

Let us now turn to thermodynamics. Using Jensen's inequality, Eq. (\ref{eq:FT1}) implies:
\begin{align}
\label{eq:2ndLaw1}
\sum_l \beta_l \langle  \Delta E_l \rangle \geq- \ln{\gamma}
\end{align}
In the case when the map is unital it is $\gamma=1$, and the second law of thermodynamics is recovered \cite{Campisi16JPA49a}. When $\gamma>1$ the condition $\sum_l \beta_l \langle  \Delta E_l \rangle <0 $ is not forbidden, and the apparent violation of the second law becomes possible. This occurs with a proper ``demonic'' design of the feedback control. When $\gamma<1$ instead the second law is more strictly enforced by means of an ``angelic'' intervention.

As shown in Refs. \cite{Morikuni11JSP143,Sagawa10PRL104} in the case of a single measurement (in either classical or quantum systems) the fluctuation relation can be restored if an information theoretic term, in the form of a mutual information, is added to the exponent in the exponential average. Ref. \cite{Horowitz10PRE82} reports the extension to the case of repeated measurements in the classical scenario. All these results are for a single-temperature initial state. In the present set-up we find as well an information theoretic correction term (see the appendix for a proof):
\begin{align}
\label{eq:FT2}
\langle e^{-\sum_l \beta_l \Delta E_l- J_{\mathbf{k},\mathbf{j}}} \rangle =1
\end{align}
where $J_{\mathbf{k},\mathbf{j}}$ is defined by the following set of equations:
\begin{align}
J_{\mathbf{k},\mathbf{j}} &= \ln \frac{p(\mathbf{k},\mathbf{j})}{p(\mathbf{j}:\mathbf{k})p(\mathbf{k})}\\
p(\mathbf{k},\mathbf{j}) &= \sum_{n,m} p(m,\mathbf{k},\mathbf{j},n)\\
p(\mathbf{k})&= \sum_{n,,\mathbf{j},m} p(m,\mathbf{k},\mathbf{j},n)= \sum_\mathbf{j} p(\mathbf{k},\mathbf{j})\\
p(\mathbf{j}:\mathbf{k}) &= \frac{p(\mathbf{k},\mathbf{j})}{\Pi_i \varepsilon[k_i | j_i]} \label{eq:pj|k}
\end{align}
The symbol $p(\mathbf{k},\mathbf{j}) $ represents the joint probability that the sequence $\mathbf{j}$ is realised and the sequence $\mathbf{k}$ is recorded, while $p(\mathbf{k})$ is the probability that $\mathbf{k}$ is recorded. The symbol $p(\mathbf{j}:\mathbf{k})$ stands for the probability that the sequence $\mathbf{j}$ is realised, conditioned on $\mathbf{k}$ being the record. 
More explicitely
\begin{align}
p(\mathbf{j}:\mathbf{k}) &= \frac{p(\mathbf{k},\mathbf{j})}{\Pi_i \varepsilon[k_i | j_i]}
=  \sum_{n,m}\Tr\,  P_m B_{\mathbf{k},\mathbf{j}} U_0  P_n U_0^\dagger B_{\mathbf{k},\mathbf{j}}^\dagger P_m p^0_n \\
B_{\mathbf{k},\mathbf{j}}&=\overleftarrow{\Pi}_i\left(U_{\mathbf{k}_i}\pi_{j_i}\right) = \frac{A_{\mathbf{k},\mathbf{j}}}{\Pi_i \sqrt{\varepsilon[k_i | j_i]}}\, . \label{eq:Bkj}
\end{align}
The operators $B_{\mathbf{k},\mathbf{j}}$ differ from the operators $A_{\mathbf{k},\mathbf{j}}$ by the term containing the conditional probability $\varepsilon[k_i | j_i]$. Note that the Bayes rule does not apply here, i.e. generally it is $p(\mathbf{k},\mathbf{j}) \neq p(\mathbf{j}:\mathbf{k})p(\mathbf{k})$. The reason is that $\mathbf{j}$ and $\mathbf{k}$ are concatenated with each other. An outcome $j_i$ influences the record $k_i$, which in turn influences the next outcome $j_{i+1}$ and so on. The quantity $J_{\mathbf{k},\mathbf{j}}$ measures the degree of such mutual influence, or correlation between the two sequences $\mathbf{j}$ and $\mathbf{k}$ \footnote{Eq. (\ref{eq:FT2}) is reminiscent of a similar relation reported by Vedral \cite{Vedral12JPA45}, see Eq. (8) there. The two relations fundamentally differ in various respects. Notably in the meaning of the mutual information term. In our case measuring the correlation between outcomes and their records, in the case of Ref. \cite{Vedral12JPA45} measuring the correlation between the measurements themselves}. In absence of feedback, namely when there is no correlation between the two sequences, $J_{\mathbf{k},\mathbf{j}}$ is null and the standard relation is recovered. Note that given a feedback rule, generally $\langle J_{\mathbf{k},\mathbf{j}} \rangle $ would grow with the length $K$ of the sequences, i.e. the number of measurements. It is accordingly expected that $\langle J_{\mathbf{k},\mathbf{j}} \rangle \propto K$ in the large $K$ regime. 

With Jensen's inequality Eq. (\ref{eq:FT2}) implies
\begin{align}
\label{eq:2ndLaw2}
\sum_l \beta_l \langle \Delta E_l  \rangle \geq - \langle J_{\mathbf{k},\mathbf{j}} \rangle
\end{align}
We thus have found two bounds to $\sum_l \beta_l \langle \Delta E_l  \rangle$.

By looking directly at the $\sum_l \beta_l \langle \Delta E_l  \rangle$ as in Ref. \cite{Campisi16JPA49a} we have found a third bound whose interpretation is  most direct and straightforward. Let
\begin{align}
\rho_\tau &=  \sum_{n,\mathbf{j},\mathbf{k},m} P_m A_{\mathbf{k},\mathbf{j}} U_0 P_n \rho_0 P_n U_0^\dagger A_{\mathbf{k},\mathbf{j}}^\dagger P_m = \sum_{\mathbf{j},\mathbf{k}}  A_{\mathbf{k},\mathbf{j}} U_0 \rho_0  U_0^\dagger A_{\mathbf{k},\mathbf{j}}^\dagger
\end{align}
be the system density matrix at time $\tau$. In the second equality we have used completeness $\sum_m P_m=\mathbb{1}$ and the fact that the initial state has no coherences in the energy eigenbasis $\sum_n P_n \rho_0 P_n = \rho_0$. Simple manipulations, similar to those employed in Ref.  \cite{Campisi16JPA49a} lead to the following salient result
\begin{align}
\sum_l \beta_l \langle \Delta E_l \rangle  = \sum_i D[\rho^l_\tau || \rho_0^l] + I[\rho_\tau] + \Delta \mathcal H
\end{align}
where 
\begin{align}
D[\rho^l_\tau || \rho_0^l] &= \Tr \rho^l_\tau \ln \rho^l_\tau - \Tr \rho^l_\tau \ln \rho^l_0 \label{eq:D}\\
I[\rho_\tau] &= -\sum_l \Tr \rho_\tau^l \ln \rho_\tau^l + \Tr \rho_\tau \ln \rho_\tau \label{eq:I}\\
\Delta \mathcal H &= -\Tr \rho_\tau \ln \rho_\tau + \Tr \rho_0 \ln \rho_0 \label{eq:DH}
\end{align}
denote the Kullback Leibler divergence between the final state $\rho_\tau$ and the initial state $\rho_0$, Eq. (\ref{eq:D});
the total amount of correlations (mutual information) that builds up among the partitions as a consequence of their interaction during the time span $[0,\tau]$, Eq. (\ref{eq:I}); and the total change in von-Neumann entropy of the whole compound, Eq.  (\ref{eq:DH}). Here $\rho_t^l= \Tr'_l \rho_t$ is the reduced state of partition $l$ at time $t$ ($\Tr'_l$ denotes trace over all partitions but the $l$-th). 
The mutual information $I$ among the partitions of the system (measuring all correlations, quantal and classical), which develops generally due to their interaction $V(t)$ (and can also occur in absence of measurements and feedback \cite{Campisi16JPA49a}), should not be confused with the classical mutual information $J_{\mathbf{k},\mathbf{j}}$ between the realisation sequence $\mathbf{j}$ and the record sequence $\mathbf{k}$ caused by the feedback mechanism.

Both the  Kullback Leibler divergence $D[\rho^i_\tau || \rho_0^i]$ and the mutual information $I[\rho_t]$ are non negative quantities. We thus arrive at the central inequality:
\begin{align}
\sum_l \beta_l \langle \Delta E_l \rangle  \geq \Delta \mathcal H 
\label{eq:2ndLaw3}
\end{align}
In the standard no measurement case, $\rho_\tau$ is linked to $\rho_0$ via a unitary map, hence $\Delta \mathcal H =0$ and one recovers the result of Ref. \cite{Campisi16JPA49a}, namely $\sum_i \beta_i \langle \Delta E_i \rangle  = \sum_i D[\rho^i_\tau || \rho_0^i] + I[\rho_\tau]$, and the second law in its standard form.
Note that when there are measurements, but no feedback, the $\rho_\tau$ is linked to $\rho_0$ via a \emph{unital} map, implying $\gamma = 0$, $\langle J_{\mathbf{k},\mathbf{j}} \rangle= 0$, and $\Delta \mathcal H \geq 0$ hence $\sum_l \beta_l \langle \Delta E_l \rangle  \geq \Delta \mathcal H \geq0$, meaning that, as is already known \cite{Campisi10PRL105,Campisi09PRE80,Watanabe14PRE89} the second law is not altered by the mere application of projective measurements that interrupt an otherwise unitary dynamics. However Eq. (\ref{eq:2ndLaw3}) clearly indicates that  there is a dissipation term associated with quantum-mechanical measurements, which is not present in the classical case. In sum through Eq. (\ref{eq:2ndLaw3}) we see that there is a thermodynamic cost associated to quantum measurements.

 Combining Eqs. (\ref{eq:2ndLaw1},\ref{eq:2ndLaw2},\ref{eq:2ndLaw3}) the second law of thermodynamics, in presence of feedback control takes the form
\begin{align}
\sum_l \beta_l \langle \Delta E_l \rangle  \geq \max[-\ln \gamma,-\langle J_{\mathbf{k},\mathbf{j}} \rangle ,\Delta \mathcal H ]
\label{eq:2ndLaw4}
\end{align}

\section{Illustrative example}
\begin{figure}[t]
\includegraphics[width=\linewidth]{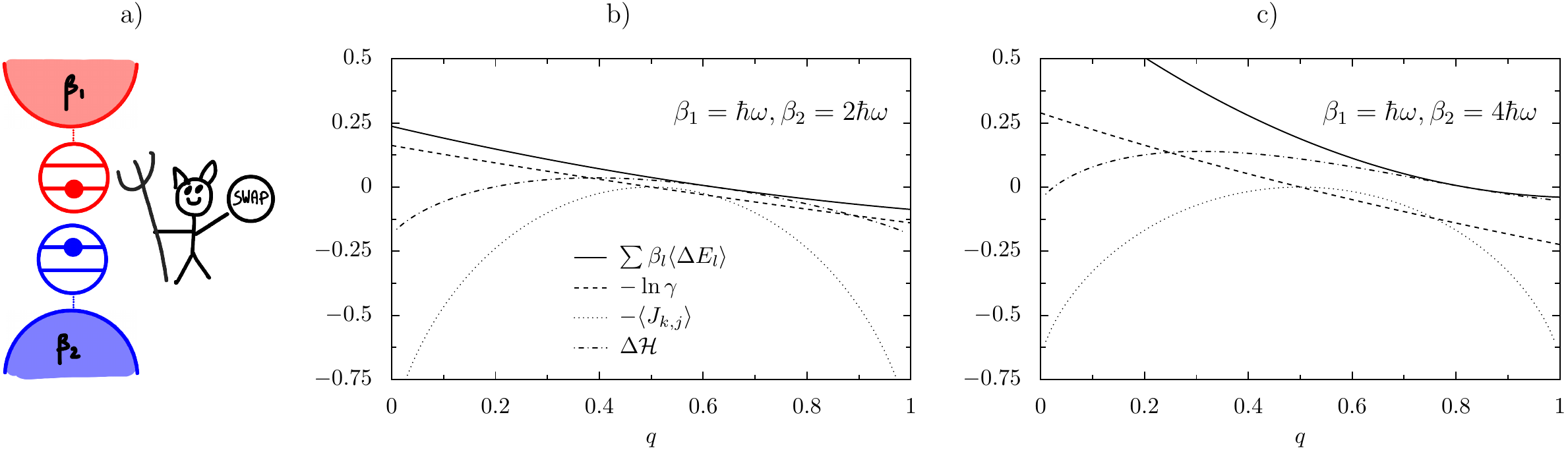}
\caption{Panel a). Scheme of a two-qubit feedback controlled refrigerator. Two qubits are prepared each in thermal equilibrium with a thermal bath. When the Demon sees the cold qubit in the excited state and the hot qubit in ground state, he swaps them. He then lets them thermalise each with its own bath and starts over. He thus transfers heat from the cold bath to the hot bath without investing energy. Panels b,c). $\sum \beta_l \langle \Delta E_l \rangle$, $-\ln \gamma$, $-\langle J_{k,j} \rangle$, $\Delta \mathcal H$ as a function of the error probability $q$ for $\beta_1=\hbar\omega$, and two different values of $\beta_2$.}
\label{fig:2}
\end{figure}

To exemplify the theory above we consider a prototypical model of quantum heat engine whose working substance is made of two qubits \cite{Campisi15NJP17,Campisi16JPA49a,Quan07PRE76}. Their Hamiltonian reads
\begin{align}
H= H_1 + H_2 = \frac{\hbar \omega}{2}\sigma_z^1 + \frac{\hbar \omega}{2}\sigma_z^2
\end{align}
where $\sigma_z^i$ denote Pauli operators. We assume the two qubits have same level spacing $\hbar\omega$ and are initially in the state:
\begin{align}
\rho_0 = \frac{e^{-\beta_1 H_1}}{Z_1} \otimes  \frac{e^{-\beta_2 H_2}}{Z_2}
\end{align}
with $Z_i$ their partition functions. At $t=0$ the $\sigma_z^i$'s are measured collapsing the two qubits in the state $|k\rangle = |k'\rangle  |k''\rangle$, with $k',k''= \pm, \pm$. We assume classical error in the measurement of each qubit $\epsilon[+|+]=\epsilon[-|-]=q$, $\epsilon[-|+]=\epsilon[+|-]=1-q$ for some $q \in [0,1]$. Accordingly the eigenvalues $j=j',j''$ are recorded with probability $\varepsilon[k|j]= \epsilon[k'|j']\epsilon[k''|j'']$. If the states $|+,+\rangle , |-,+\rangle, |-,-\rangle$, are recorded we do nothing: $U_{+,+}= U_{-,+}= U_{-,-}=\mathbb{1}$; else, i.e., if $k=|-,+\rangle $ we apply a swap operation, $U_{-,+}=U_{SWAP}$, that maps $|-,+\rangle$ into  $|+,-\rangle$. The system is now in a joint eigenstate $|m\rangle= U_k |j\rangle$ of the two qubits Hamiltonian $H$, hence the final measurement of the $\sigma_z^i$ is irrelevant. At the end of the process each qubit is allowed to relax to thermal equilibrium with their respective thermal baths of inverse temperatures $\beta_i$ so as to re-establish the initial state $\rho_0$. Accordingly the average energies $\langle \Delta E_i \rangle $ acquired by each qubit during the process equals the average heats that they release in the baths in the thermal relaxation step. Due to the feedback mechanism energy may be withdrawn from the cold bath and released in the hot one. Note that, due to the fact that the two qubits have same level spacing the SWAP operation does not alter their total energy. Namely there is no energy injection by the Demon: to steer the energy flow he only uses information. The set-up is illustrated in Fig. \ref{fig:2} panel a).

The relevant probability  chain  is a bit simpler than in the general case because  the first energy measurement is itself here also the first feedback measurement. It reads $
p(m,k,j) = \Tr P_m U_k P_j U_k^\dagger P_m p_j^0 \varepsilon[k|j] = \Tr P_m A_{k,j} A_{k,j}^\dagger P_m p_j^0$ with $
A_{k,j} = \sqrt{\varepsilon[k|j]} U_k P_j $. For $\gamma$ we have $\gamma= \sum_{j,k} {\varepsilon[k|j]} \Tr P_j U_k^\dagger \rho_0 U_k P_j $. The final state is $\rho_f=  \sum_{j,k} {\varepsilon[k|j]} U_k P_j \rho_0  P_j U_k^\dagger$. The probability $p(j:k)$ that the outcome $j$ is realised conditioned on $k$ being recorded is simply the marginal probability  $p(j)$ that $j$ is realised because the record $k$ comes chronologically after the realisation of $j$ and hence cannot have any influence on it. The quantity $J_{k,j}$ boils down then to the logarithm of the ratio $p(j,k)/p(j)p(k)$ \cite{Morikuni11JSP143} hence its expectation is the non-negative mutual information between $j$ and $k$: $\langle J_{k,j} \rangle=\sum_{j,k} p(j,k) [\ln p(j,k)/p(j)p(k)]$.

Panels b,c) of Fig. \ref{fig:2} show $\sum \beta_l \langle \Delta E_l \rangle, -\ln \gamma, -\langle J_{k,j}\rangle , \Delta \mathcal H$ for two choices of $\beta_2$ and same $\beta_1$, as a function of the error probability $q$. 
 In accordance with Eq. (\ref{eq:2ndLaw4}) we see that $\sum \beta_l \langle \Delta E_l \rangle$ is bounded from below by $-\ln \gamma, -\langle J_{k,j}\rangle$ and  $\Delta \mathcal H$. Independent of all other parameters the refrigerator cannot work in the region $q<1/2$ where $j$ and $k$ are anti-correlated, while it may only work if $q>1/2$. This is captured by $-\ln \gamma$ being positive in the region $0<q<1/2$ and negative for $1/2<q<1$. At $q=1/2$ outcome and recording are fully uncorrelated, which restores unitality as discussed above and implies $\ln \gamma=0$. Regarding $\Delta \mathcal H$, while it tends to be closer to $\sum \beta_l \langle \Delta E_l \rangle$ in the operation region ($q>1/2$), it greatly departs from it in the non-operation region, where it can even get negative values. Notably in both panels there is a value of $q$ for which the bound is saturated by $\Delta \mathcal H$. Regarding  $-\langle J_{k,j}\rangle$ we note it is everywhere non-positive as expected. Furthermore it is symmetric with respect to $q\rightarrow 1-q$. This reflects the fact that the mutual information does not distinguish between correlation and anti-correlation. The maximum $-\langle J_{k,j}\rangle =0$ is attained at $q=1/2$ where $j,k$ are uncorrelated, and the standard fluctuation relation is recovered (i.e., $\gamma=1$). In both panels we see that $\Delta \mathcal H > -\langle J_{k,j}\rangle$. Whether this a generic bound is yet to be understood. We note that while at $q=1/2$ both $-\ln \gamma$ and $-\langle J_{k,j}\rangle$ are null, $\Delta \mathcal H$ is non-negative, reflecting the fact that in absence of feedback there is nonetheless an entropic cost associated to measurements, as discussed above. Such cost can be counterbalanced in presence of feedback (note that $\Delta H$ may be negative for $q \neq 1/2$).  Confronting now the two panels, we see that the higher the thermal gradient $\beta_2-\beta_1$, the larger is the point $q$ where the engine starts operating, i.e. where $\sum \beta_l \langle \Delta E_l \rangle$ turns from positive into negative: As intuition suggests the more the gradient the better must your measurement be. This feature is captured also by $\Delta \mathcal H$ but not by $-\ln \gamma, -\langle J_{k,j}\rangle$. Also the smaller the gradient the more the shape of the function $\sum \beta_l \langle \Delta E_l \rangle$ resembles that of $-\ln \gamma$, with the shift between the two being approximately the value of $\Delta \mathcal H$ at $q=1/2$: that is $\sum \beta_l \langle \Delta E_l \rangle \simeq -\ln \gamma + \Delta \mathcal H|_{q=1/2}$.

\section{Experimental proposal}

\begin{figure}[b]
\includegraphics[width=\linewidth]{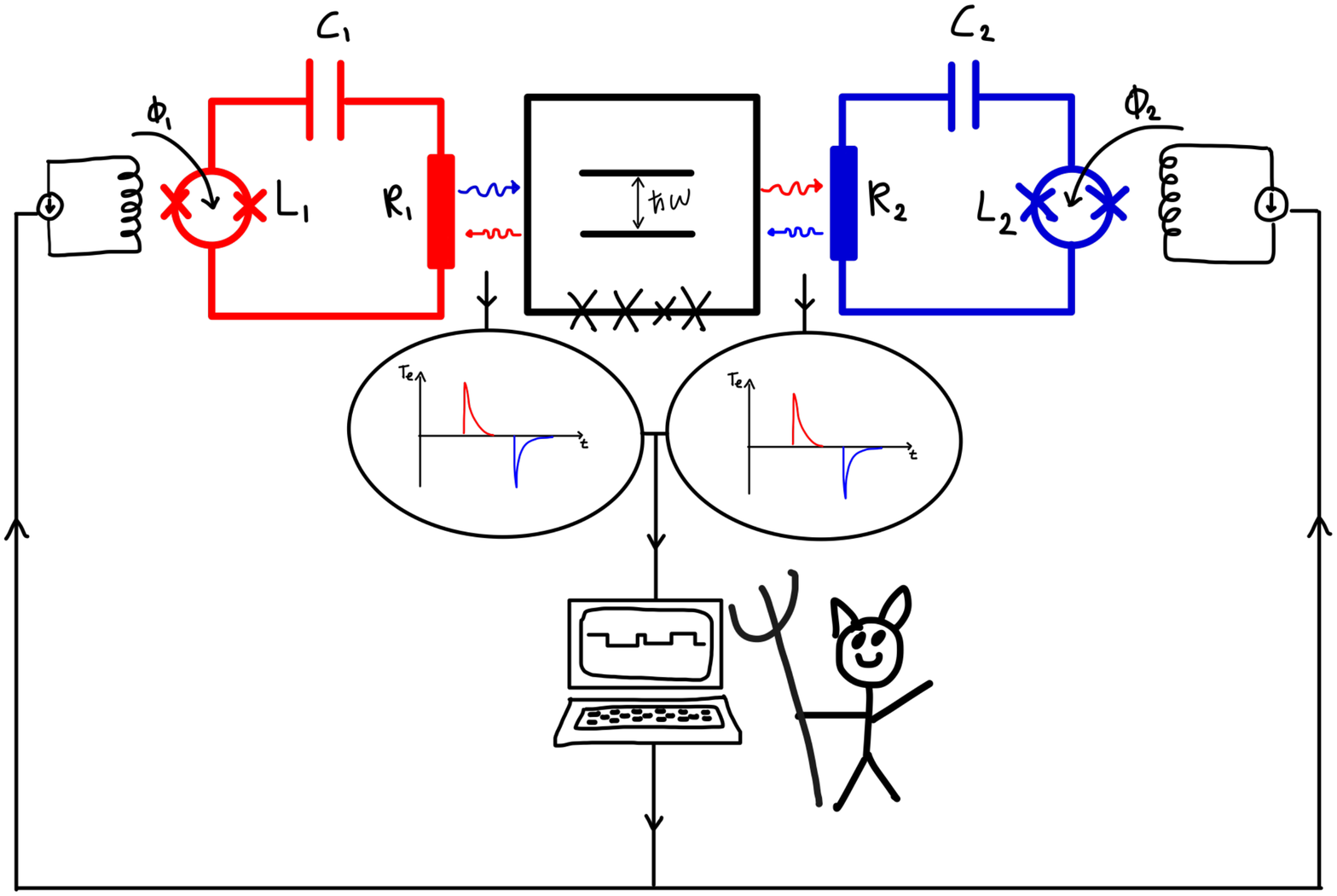}
\caption{Set up of the proposed experiment. A superconducting qubit (black rectangle) embodies a Maxwell demon trap door, and two resistors embedded in RLC circuits embody the two chambers of different temperatures. Qubit and RLC circuits are inductively coupled. Calorimetric monitoring of photons entering and exiting each resistor is applied, allowing to both measure heat exchanged by each resistor, and monitoring the state of the qubit at any time. When the qubit is up a feedback algorithm drives the resonance frequency of the cold RLC circuit out of tune with the qubit frequency, while keeping the hot RLC in tune with it (and vice versa) so that an overall heat current flows from cold to hot. The resonance frequencies of the RLC circuits are controlled by tuning their non-linear inductive elements, i.e., SQUIDs, via application of external magnetic flux $\Phi_i$. }
\label{fig:3}
\end{figure}
The general theory developed above allows for a joint information theoretic and  thermodynamic analysis of feedback controlled dynamics in the broad scenario where 
a demon can influence not only the amount of work being provided by the outside as in previous works  \cite{Toyabe10NP6,Koski14PRL113,Koski14PNAS111}, but also the heat flow between the various parts of a compound system, e.g. the heat flow between various heat baths. 

The progress of solid state technology on the other hand allows to realise such feedback controlled heat transport mechanisms in real devices. The example illustrated above can be experimentally realised by introducing a feedback mechanism in the two-superconducting qubits scheme illustrated in in Ref. \cite{Campisi15NJP17}. Below we illustrate a design that is of more immediate realisation. It is a based on a single qubit and it does not involve any qubit-operation, but only manipulations of qubit-bath couplings. The proposal that we put forward here is based on two ingredients that enable unique capabilities allowing for the implementation of a Maxwell demon based on a most simple concept. The two ingredients are a two-level-system acting as quantum trap door and the calorimetric measurement scheme developed in Refs. \cite{Pekola10PRL105,Gasparinetti15PRAPP3}. 

 The one qubit set-up is illustrated in Fig. \ref{fig:3}. The two-level system is embodied by a superconducting qubit of level spacing $\hbar \omega$. The two chambers are embodied by two resistors being kept at different temperatures. Qubit and resistors con exchange energy (i.e. heat) in the form of photons of energy $\hbar \omega$ associated to the TLS absorbing/emitting one photon from/to one of the two baths. The resistors are embedded into an RLC loop of tunable resonance frequency. This results into a tuneable TLS/resistor coupling. When an RLC circuit is far detuned from $\omega$, the qubit is effectively decoupled from the resistor, while maximal coupling occurs when it is in tune with the qubit. The resonance frequency can be tuned by using a SQUID as a non-linear and tuneable inductor, its inductance being governed by a controllable threading magnetic flux.

When a photon enters/exits one of the two resistors, its electronic temperature undergoes a positive/negative jump followed by a fast decay. Two calorimeters \cite{Pekola10PRL105,Gasparinetti15PRAPP3} continuously monitor the two resistors, and count how many photons enter/exit them. This allows for a directional full counting statistics of heat. Most remarkably it also allows to infer the state of the TLS at each time. If an absorption (in either resistor) is observed,
it means the TLS jumped down, hence it was up before the absorption was detected, and is down afterwards. This allows to experimentally access the quantum state trajectory of the TLS.

The feedback concept is extremely simple: as soon as a jump-down is observed, turn on the interaction with the cold resistor and turn off the interaction with the hot resistor. Vice-versa for the observation of a jump up. This results in a net flow of heat from the cold resistor to the hot one. Based on the above general analysis
the apparent violation of the second law is understood in terms of lack of time-reversal symmetry of feedback control, leading to an overall non-unital dynamics of resistors plus TLS. In a practical realisation one is realistically not able to fully turn off the interactions.
Furthermore there will be some delay time $\delta$ between measurement being performed and feedback being realised, giving rise effectively to possible error $\varepsilon[k_i|j_i]$ between measured state $k_i$ and actual state $j_i$ of the qubit.

\section{Modelling}
In the following we model the dynamics of the proposed experiment.
We model the evolution of the two level system via a standard Lindblad master equation
\begin{equation}
\dot \rho = -i [H_S,\rho] + \mathcal L_L \rho + \mathcal L_R \rho
\end{equation}
where $H_S= -E_0 (\Delta \sigma_x + q \sigma_z)=$ is the two level system Hamiltonian expressed in terms of the Pauli matrices $\sigma_\alpha$, and $\mathcal{L}_l$ are Lindblad operators
\begin{align}
\mathcal L_l \rho &= \Gamma_l^\downarrow D[\sigma] \rho + \Gamma_l^\uparrow D[\sigma^\dagger]\rho
\end{align}
expressed in terms of and the super-operator $D[O]\rho = O\rho O^\dagger -\frac{1}{2} O^\dagger O \rho- \frac{1}{2}\rho O^\dagger O $ and the rising and lowering spin operators $\sigma^\dagger,\sigma$ of the Hamiltonian $H_S$, defined via $\sigma^\dagger |-\rangle = |+\rangle, \sigma^\dagger |+\rangle = 0, \sigma |+\rangle=  |-\rangle, \sigma |-\rangle=0$, where $|- (+)\rangle$ is the ground (excited) state of $H_S$. Here $l=L,R$ denote either the left or the right reservoir.
The rates $\Gamma_l^{\downarrow\uparrow}$ for jump down/up in the $l$'th resistor are given by
\begin{align}
\label{eq:rates}
\Gamma_l^{\downarrow\uparrow} &= \frac{E_0^2 M_l^2}{\hbar^2 \Phi_0}  \frac{\Delta^2}{(q^2+\Delta^2)} S_{I,l}(\pm \omega)
\end{align}
where
$
S_{I,l}(\omega) = S_{V,l}(\omega) [R_l^2(1+Q_l^2[\omega/\omega_{LC,l}-\omega_{LC,l}/\omega ]^2)]^{-1}
$
is the current noise spectrum expressed in terms of the voltage noise spectrum 
$
S_{V,l}(\omega) =2 R_l \hbar \omega (1-e^{-\beta_l \hbar \omega})^{-1}
$,
$Q_l=\sqrt{L_l / C_l}/R_l $ is the quality factor and $\omega_{LC,l}=1/\sqrt{L_l C_l}$ the resonance frequency of resonator $l$,
expressed in terms of its resistance, inductance and capacitance $R_l,L_l,C_l$. By increasing $L_j$ the rates $\Gamma_l^{\downarrow\uparrow}   $, can be quenched, namely the interaction between the TLS and the $l$-th resistor can be turned off. 
The symbol $M_l$ stands for the mutual inductance between the qubit and the $l$-th resistor and $\Phi_0$ is the flux quantum. 
Note that the rates are detailed balanced:
\begin{align}
\Gamma_l^{\downarrow} = e^{\beta_l \hbar \omega} \Gamma_l^{\uparrow}
\end{align}
The study of heat and work fluctuations requires the study of the dynamics to be performed at the level of single quantum-jump trajectories \cite{Campisi15NJP17,Hekking13PRL111}, resulting from the unravelling of the master equation. This is here achieved by means of the Monte Carlo wave function (MCWF) method  \cite{Molmer93JOSAB,BreuerPetruccioneBOOK}. In the specific case under study of a two level system subject to dissipation terms leading to full wave function collapse in either state $|-\rangle$ or $|+\rangle$, this results in a classical dichotomous Poisson process with rates $\Gamma_l^{\downarrow\uparrow}$ \cite{Campisi15NJP17}. 

The basis of our numerical experiment is the generation of such dichotomous Poisson random trajectories. We chose the right reservoir as the cold one and the left as the hot one. The TLS is assumed to be initially in equilibrium with the left bath. We produce a large sample of trajectories and build the normalised historgram $h(N_R)$ of the number $N_R$ of photons entering the right reservoir. Since the heat $Q_R$ entering the right reservoir is given as $Q_R=\hbar N_R \omega$, the statistics $h(N_R)$ is the heat statistics. In absence of feedback it satisfies the fluctuation relation 
\begin{align}
\label{eq:heatFT}
\frac{h(N_R)}{h(-N_R)}= e^{-\Delta \beta \hbar\omega N_R}, \qquad \text{no feedback}
\end{align}
The feedback is introduced as follows. At each moment in time we distinguish between the actual state of the system $j=\pm$ and the knowledge $k=\pm$ we have about it. The latter does not necessarily coincide with the former because we allow for some delay-time $\delta$ between a jump occurring in the TLS and our knowledge of the state of the qubit being updated accordingly. The delay time thus effectively introduces an error probability $\varepsilon[\pm|\pm]$ between the 
actual state and the knowledge about the state, at each time. At each time, conditioned on the knowledge $k$ of the state we use either one set of rates favouring the interaction with either the cold or hot bath. More explicitly, let $\Gamma^{\downarrow\uparrow|\pm}_{l}$ be the rate for jump down (up) in $l$-th bath conditioned on TLS being measured to be in state $\pm$. In accordance with Eq. (\ref{eq:rates}) we use the following rates
\begin{align}
\Gamma^{\downarrow|+}_{L}= \frac{A}{1-e^{-\beta_L \hbar \omega}} \qquad \Gamma^{\uparrow|+}_{L}= \Gamma^{\downarrow|+}_{L} e^{-\beta_L \hbar \omega} \\
\Gamma^{\downarrow|+}_{R}= \frac{B}{1-e^{-\beta_R \hbar \omega}} \qquad \Gamma^{\uparrow|+}_{R}= \Gamma^{\downarrow|+}_{R}e^{-\beta_R \hbar \omega}\\
\Gamma^{\downarrow|-}_{L}= \frac{B}{1-e^{-\beta_L \hbar \omega}} \qquad \Gamma^{\uparrow|-}_{L}= \Gamma^{\downarrow|-}_{L} e^{-\beta_L \hbar \omega} \\
\Gamma^{\downarrow|-}_{R}= \frac{A}{1-e^{-\beta_R \hbar \omega}} \qquad \Gamma^{\uparrow|-}_{R}= \Gamma^{\downarrow|-}_{R}e^{-\beta_R \hbar \omega}
\end{align}
where $A,B$ are determined by the circuitry parameters, and can be tuned via external fluxes $\Phi_i$. With $B<A$, this means that energy exchange with the right (cold) bath is larger when the TLS is believed to be down, so that it becomes more likely that energy flows out of the cold reservoir. Similarly energy exchange with the left (hot) bath is larger when the TLS is believed to be  up, so that it becomes more likely that energy flows in the hot reservoir.
Overall this results in an effect that contrasts the natural flow from hot to cold. The largest effect can be achieved when turning off the unwanted interaction completely, namely when $B=0$. Having in mind a realistic set-up here we keep the ratio $A/B$ finite, meaning partial turning-off is considered.
\begin{figure}[t]
\includegraphics[width=\linewidth]{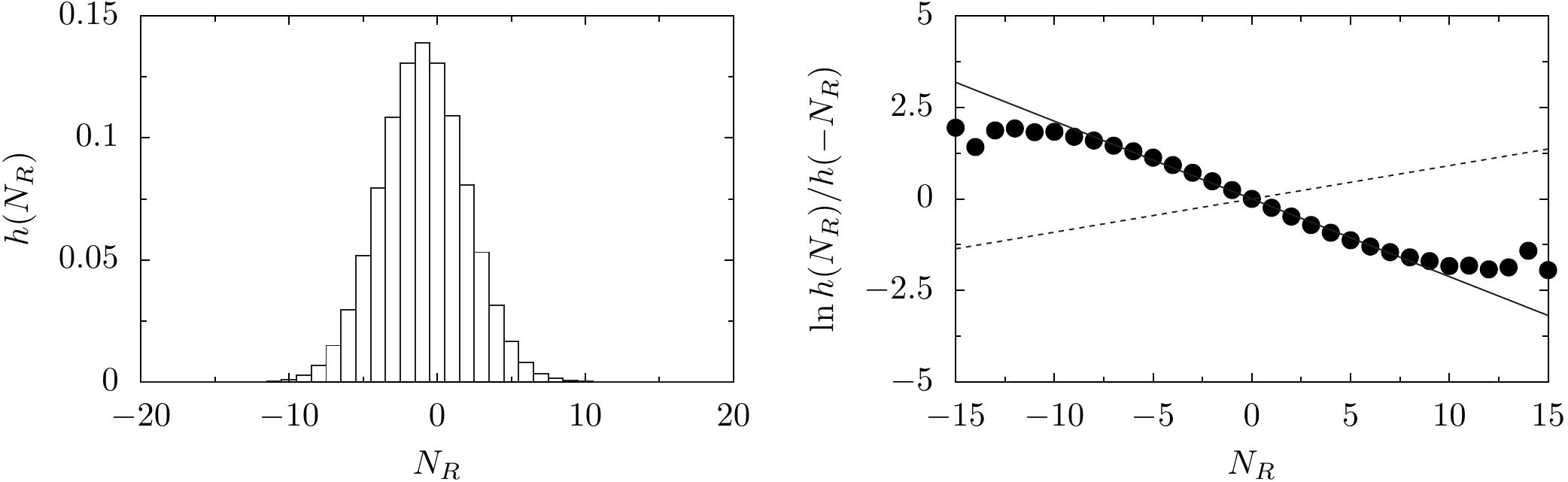}
\caption{Left: typical histogram $h(N_R)$. Right: $\ln h(N_R)/h(-N_R)$ as a function of $N_R$. Straight dashed line is $\Delta \beta \hbar \omega N_R$.  Straight solid line is $-\Delta \beta_\text{eff} \hbar \omega N_R$. Here $k_B T_L = 1.1,k_B T_R=1$. These thermal energies are expressed in units of $-\hbar \omega$. $\omega$ also fixes the time unit.  Delay time is $\delta=0.5\times 10^3$ in those time units. It is $A=1 \times10^{-3}$, $B=0.5 \times 10^{-3}$ corresponding to the largest rate timescale $\bar t =1.2642 \times 10^3$. The simulation time is $20 \bar t$. The statistics is built on a sample of $5 \times 10^6$ trajectories. }
\label{fig:4}
\end{figure}

Because of the feedback the fluctuation relation (\ref{eq:heatFT}) is not obeyed. However it can be proved (see appendix) that, due to the feedback mechanism, the TLS feels the effective temperature gradient 
\begin{align} 
\label{eq:DeltaTeffective}
\Delta \beta^\text{eff}=\beta^\text{eff}_L- \beta^\text{eff}_R = \Delta \beta + \frac{2}{\hbar\omega}\ln \frac{\varepsilon[+|+] A+ {\varepsilon[-|+] B}}{{\varepsilon[+|-] A+\varepsilon[-|-]B}} 
\end{align}
we thus see that by tuning the ratio $A/B$ the effective temperature gradient can be manipulated and if the errors associated to the measurement is not too big, it can even be inverted as compared to the original thermal gradient $\Delta \beta$.
So the overall effect of the demon is to change the ``temperatures felt'' by the TLS. Accordingly the following fluctuation relation
\begin{align}
\label{eq:heatFTeffective}
\frac{h(N_R)}{h(-N_R)}= e^{-\Delta \beta_\text{eff} \hbar\omega N_R}
\end{align}
is obeyed by the histogram $h(N_R)$. This immediately allows to interpret the quantity
\begin{align}
J_\text{exp}=-\frac{2 Q_R}{\hbar\omega}\ln \frac{\varepsilon[+|+] A+ {\varepsilon[-|+] B}}{{\varepsilon[+|-] A+\varepsilon[-|-]B}} 
\end{align}
via Eq. (\ref{eq:FT2}) as the mutual information encoded in a trajectory along which a heat $Q_R$ is exchanged with the $R$ bath. Note that when $A=B$, the feedback has no effect and accordingly $J_\text{exp}=0$. Likewise if $\varepsilon[+|+]=\varepsilon[+|-]$ (hence $\varepsilon[-|+]=\varepsilon[-|-]$) meaning no correlation between state and knowledge thereof, feedback control does not work and again $J_\text{exp}=0$. Most importantly the experimental mutual information $J_\text{exp}$ is proportional to the heat exchanged. This allows for accessing a fluctuating information theoretic quantity by means of a thermodynamic measurements in a realistic experimental scenario. 

Figure \ref{fig:4} shows typical histograms $h(N_R)$ for realistic parameters. We also plotted the quantity $\ln h(N_R)/h(-N_R)$ finding a good agreement with the theoretical prediction $-\Delta \beta_\text{eff} \hbar\omega N_R$. The effective conditional probabilities $\varepsilon[k|j]$ were obtained by recording for each trajectory the total time when state was $j$ and knowledge was $k$, and averaging their value over the whole ensemble of trajectories. The observed deviation is a consequence of the fact that error here is not introduced in the form of an outcome being missed (as assumed in deriving Eq. (\ref{eq:heatFTeffective})), but rather being reported with some delay. 
With the histogram $h(N_R)$ we computed $\sum \beta_l \langle \Delta E_l\rangle = \hbar \omega \Delta \beta \langle N_R \rangle = -0.0862$, $-\ln \gamma = -\ln \langle e^{\Delta \beta \hbar \omega N_R } \rangle=-0.1205$, $-\langle J_\text{exp}\rangle= -0.2873 $, for the chosen parameters. The computed values are in agreement with the prediction of Eq. (\ref{eq:2ndLaw4}). The proposed experiment does not allow to measure $\Delta \mathcal H$, which would require accessing the full system+baths density matrix.

\subsection{Energy spent by the Demon}
What is the energy cost incurred by the demon to open/close the trap-door? To roughly estimate that we model the LCR circuit as a classical harmonic oscillator (LC circuit) in contact with a heat bath (the resistor) at temperature $T$. To open/close the door towards one of the two reservoirs, the demon switches the LC frequency from $\omega_i$ to another frequency $\omega_f$ so as to put it in/off resonance with the qubit. If the operation is carried in a quasi static manner, the work done  is equal to the free energy change: $W= k_B T \ln (\omega_f/\omega_i)$. The operation would in this case be reversible, and the work lost when opening the door will be retrieved when opening it. The overall cost of a open/close cycle would be null in this limiting case. The other limiting case is when the switch is infinitely fast. The overall cost of a single open/close cycle in this case would be non-negative in accordance with the second law of thermodynamics, and amounts to $W= k_B T (\omega_f/\omega_i- \omega_i/\omega_f)^2/2$. The overall work incurred in a repeated feedback operation is proportional to the number of open/close cycles, which in turn is proportional to the net number of energy quanta being transported, namely the total heat transported. Interestingly we note that the faster the open/close operation, the more effective is the feedback mechanism, the more energy needs to be invested. 

\section{Conclusions}
We have developed a general quantum theory of repeated feedback control in a multiple heat reservoir scenario. The main effect of feedback control is that it induces a generally non-unital dynamics of the full reservoirs+system compound. As a consequence the standard bound set by the second law od thermodynamics on the dissipation quantifier $\sum_l \beta_l \langle \Delta  E_l \rangle$ is shifted and may become negative. We have illustrated an experimental proposal where  a single superconducting qubit plays the role of a trap-door that is subject to feedback control. The envisaged method for simultaneously measuring the qubit state and the heat exchanged by each reservoir is single photon calorimetry.

\section*{Acknowledgements}
This research was supported by a Marie Curie Intra European Fellowship within the 7th European Community Framework Programme through the project NeQuFlux grant n. 623085 (M.C.), by Unicredit Bank (M.C.),  by the Academy of Finland contract no. 272218 (J.P.), and by the COST action MP1209 ``Thermodynamics in the quantum regime''. 

\appendix
\section{Derivation of Eq. (\ref{eq:FT1})}
\begin{align}
\langle e^{-\sum_l \beta_l \Delta E_l} \rangle &= \sum_{n,\mathbf{j},\mathbf{k},m} p(m,\mathbf{k},\mathbf{j},n) e^{-\sum \beta_l  E_m^l}   e^{\sum \beta_l E_n^l} \nonumber \\
&= \sum_{n,\mathbf{j},\mathbf{k},m} \Tr\,  P_m A_{\mathbf{k},\mathbf{j}} U_0  P_n U_0^\dagger A_{\mathbf{k},\mathbf{j}}^\dagger P_m  \frac{ e^{-\sum \beta_l E_n^l}}{Z} e^{-\sum \beta_l  E_m^l}   e^{\sum \beta_l E_n^l} \nonumber \\
&= \sum_{\mathbf{j},\mathbf{k},m} \Tr\,  P_m A_{\mathbf{k},\mathbf{j}} A_{\mathbf{k},\mathbf{j}}^\dagger P_m  \frac{ e^{-\sum \beta_l E_m^l}}{Z} \nonumber \\
&= \sum_{\mathbf{j},\mathbf{k}}   \Tr\,   A_{\mathbf{k},\mathbf{j}}^\dagger \rho_0 A_{\mathbf{k},\mathbf{j}} \nonumber
 \end{align}
Eq. (\ref{eq:pnjkm}) and $\rho_0= \sum _n P_n   e^{-\sum \beta_l E_n^l}/Z$ have been used to obtain the second line. Completeness $\sum_n P_n= \mathbb{1}$ and unitarity $U_0  U_0^\dagger = \mathbb{1}$ led to the third line. Fourth line follows from the cyclical property of the trace, idempotence $P_mP_m= P_m$ and $\rho_0= \sum _m P_m  { e^{-\sum \beta_l E_m^l}}/{Z}$.

\section{Derivation of Eq. (\ref{eq:FT2})}
Using Eq. (\ref{eq:pj|k}), the exponentiated fluctuating mutual information can be conveniently expressed as
\begin{align}
e^{-J_{\mathbf{k},\mathbf{j}}} = \frac{p(\mathbf{j}:\mathbf{k})p(\mathbf{k})}{p(\mathbf{k},\mathbf{j})}
= \frac{p(\mathbf{k})}{\Pi_i \varepsilon[k_i | j_i]}
\end{align}
hence 
\begin{align}
\langle e^{-\sum_l \beta_l \Delta E_l- J_{\mathbf{k},\mathbf{j}}} \rangle &= \sum_{n,\mathbf{j},\mathbf{k},m} p(m,\mathbf{k},\mathbf{j},n) e^{-\sum \beta_l  E_m^l}   e^{\sum \beta_l E_n^l} \frac{p(\mathbf{k})}{\Pi_i \varepsilon[k_i | j_i]}\\
&= \sum_{n,\mathbf{j},\mathbf{k},m} \Tr\,  P_m B_{\mathbf{k},\mathbf{j}} U_0  P_n U_0^\dagger B_{\mathbf{k},\mathbf{j}}^\dagger P_m  \frac{ e^{-\sum \beta_l E_n^l}}{\Pi_l Z_l} e^{-\sum \beta_l  E_m^l}   e^{\sum \beta_l E_n^l} p(\mathbf{k})\nonumber \\
&= \sum_{\mathbf{j},\mathbf{k},m} \Tr\,  P_m B_{\mathbf{k},\mathbf{j}} B_{\mathbf{k},\mathbf{j}}^\dagger P_m  \frac{ e^{-\sum \beta_l E_m^l}}{\Pi_l Z_l}p(\mathbf{k}) \nonumber \\
&= \sum_{\mathbf{k},m} \Tr\,  P_m  \frac{ e^{-\sum \beta_l E_m^l}}{\Pi_l Z_l}p(\mathbf{k}) \nonumber \\
&= \Tr \rho_0 \sum_\mathbf{k} p(\mathbf{k})\\
&= 1
\end{align}
Eq. (\ref{eq:pnjkm}), $\rho_0= \sum _n P_n   e^{-\sum \beta_l E_n^l}/\Pi_l Z_l$ and Eq. (\ref{eq:Bkj}) have been used to obtain the second line. Completeness $\sum_n P_n= \mathbb{1}$ and unitarity $U_0  U_0^\dagger = \mathbb{1}$ led to the third line. The fourth line follows from $\sum_{\mathbf{j}} B_{\mathbf{k},\mathbf{j}} B_{\mathbf{k},\mathbf{j}}^\dagger = \mathbb{1}$ which follows by expanding the $i$-ordered products, apply idempotence $\pi_{j}\pi_{j}=\pi_{j}$, completeness $\sum_{j} \pi_{j}= \mathbb{1}$, and unitarity $U_j U_j^\dagger = \mathbb{1}$. Cyclical property of the trace, idempotence $P_mP_m= P_m$ and $\rho_0= \sum _m P_m  { e^{-\sum \beta_l E_m^l}}/{\Pi_l Z_l}$ lead to the fifth line. The final result is a consequence of normalisation of $\rho_0$ and of $p(\mathbf{k})$.

\section{Derivation of Eq. (\ref{eq:DeltaTeffective})}
Under the operation of the demon the TLS experiences effective temperatures of the baths that differ from their actual value. 
To fix ideas, let us for the moment, assume no delay time and no error in the measurement. The qubit is effectively subject to the following effective rates
$\Gamma^{\downarrow,\text{eff}}_{s} = \Gamma^{\downarrow|+}_{s}, \Gamma^{\uparrow,\text{eff}}_{s}  = \Gamma^{\uparrow|-}_{s}$.
Accordingly, the detailed balance temperatures are shifted: 
\begin{align}
 e^{\beta^\text{eff}_L \hbar \omega} &= \Gamma^{\downarrow,\text{eff}}_{L}/ \Gamma^{\uparrow,\text{eff}}_{L} = (A/B) e^{\beta_L\hbar \omega}\\
 e^{\beta^\text{eff}_R \hbar \omega} &= \Gamma^{\downarrow,\text{eff}}_{R}/ \Gamma^{\uparrow,\text{eff}}_{R} = (B/A) e^{\beta_R\hbar \omega}
\end{align}
where we used the explicit expressions Eq. (\ref{eq:rates}). This 
implies the effective temperatures
\begin{align}
  \beta^\text{eff}_L &=  \frac{1}{\hbar \omega}\ln (A/B) +\beta_L\\
  \beta^\text{eff}_R &= \frac{1}{\hbar \omega}\ln (B/A) +\beta_R
 \end{align}

Let us now introduce the errors $\varepsilon[\pm|\pm]$ related to the measurement. The stochastic process describing the dynamics of the TLS is still Poissonian with one rate occurring in case of right measurement and one rate occurring in the other case. The idea is that monitoring is continuous, or better, occurring with a sampling time interval $dt$, which we assume short compared to all rates $\Gamma_{R,L}^{\uparrow,\downarrow|\pm}$. Let us imagine the system is in state $j=+$. There is a probability $\varepsilon[+|+]$ the observation is $k=+$ and a probability $\varepsilon[-|+]$ the observation is $k=-$. Thus the probability to undergo a jump down in the $s$ reservoir in the interval $dt$ is 
\begin{align}
p(dt) &= \varepsilon[+|+] e^{-\Gamma^{\downarrow|+}_{s}dt} +  \varepsilon[-|+]  e^{-\Gamma^{\downarrow|-}_{s}dt} \\
& \simeq \varepsilon[+|+](1-\Gamma^{\downarrow|+}_{s}dt) + \varepsilon[-|+](1-\Gamma^{\downarrow|-}_{s}dt) \\
& = 1- (\varepsilon[+|+]\Gamma^{\downarrow|+}_{s}+\varepsilon[-|+]\Gamma^{\downarrow|-}_{s})dt \\
& = e^{-(\varepsilon[+|+]\Gamma^{\downarrow|+}_{s}+\varepsilon[-|+]\Gamma^{\downarrow|-}_{s})dt}
\end{align}
Similarly for the jump up. Overall the TLS experience the new rates
\begin{align}
\Gamma^{\downarrow}_{s,\text{eff}} &= \varepsilon[+|+]\Gamma^{\downarrow|+}_{s}+ \varepsilon[-|+]\Gamma^{\downarrow|-}_{s}
\\
\Gamma^{\uparrow}_{s,\text{eff}}  &=  \varepsilon[+|-] \Gamma^{\uparrow|+}_{s} +\varepsilon[-|-] \Gamma^{\uparrow|-}_{s}
\end{align}
Accordingly 
\begin{align}
 e^{\beta^\text{eff}_s\hbar \omega} &= \frac{\Gamma^{\downarrow}_{s,\text{eff}} }{\Gamma^{\uparrow}_{s,\text{eff}} } = 
 \frac{\varepsilon[+|+]\Gamma^{\downarrow|+}_{s}+ \varepsilon[-|+]\Gamma^{\downarrow|-}_{s}}{\varepsilon[+|-] \Gamma^{\uparrow|+}_{s} +\varepsilon[-|-] \Gamma^{\uparrow|-}_{s}}\\
\beta_s^\text{eff} &= \frac{1}{\hbar \omega} \ln  \frac{\varepsilon[+|+]\Gamma^{\downarrow|+}_{s}+ \varepsilon[-|+]\Gamma^{\downarrow|-}_{s}}{\varepsilon[+|-] \Gamma^{\uparrow|+}_{s} +\varepsilon[-|-] \Gamma^{\uparrow|-}_{s}}
 \end{align}
Plugging in the explicit expressions we get
\begin{align}
\beta_L^\text{eff} & = \beta_L + \frac{1}{\hbar \omega} \ln  
 \frac{\varepsilon[+|+]A+ \varepsilon[-|+]B}{\varepsilon[+|-]A +\varepsilon[-|-]  B}\\
\beta_R^\text{eff} &= \beta_R + \frac{1}{\hbar \omega} \ln  \frac{\varepsilon[+|-]A +\varepsilon[-|-]  B}{\varepsilon[+|+]A+ \varepsilon[-|+]B}
\end{align}
Hence Eq. (\ref{eq:DeltaTeffective}).

\section*{References}
\bibliographystyle{iopart-num}
\providecommand{\newblock}{}

\end{document}